\documentclass[11pt]{amsart}

\usepackage{amsfonts}
\usepackage{amsmath}
\usepackage{amssymb}

\theoremstyle{plain}
\newtheorem{theorem}{Theorem}[section]
\newtheorem{lemma}[theorem]{Lemma}

\theoremstyle{definition}

\theoremstyle{remark}

\numberwithin{equation}{section}

\begin{document}
\thispagestyle{empty}
\begin{center}
{\footnotesize Available at: 
{\tt http://www.ictp.it/\~{}pub$_-$off}}\hfill IC/2007/028\\
\vspace{1cm}
United Nations Educational, Scientific and Cultural Organization\\
and\\
International Atomic Energy Agency\\
\medskip
THE ABDUS SALAM INTERNATIONAL CENTRE FOR THEORETICAL PHYSICS\\
\vspace{2.5cm}
{\bf A BORSUK-ULAM TYPE GENERALIZATION OF THE\\
LERAY-SCHAUDER FIXED POINT THEOREM}\\
\vspace{2cm}
Anatoliy K. Prykarpatsky\footnote{prykanat@cybergal.com; pryk.anat@ua.fm}\\
{\it The Department of Nonlinear Mathematical Analysis at the IAPMM of
NAS,\\ Lviv, 79060, Ukraine,\\ 
The AGH University of Science and Technology,
Krakow 30059, Poland\\
and\\
The Abdus Salam International Centre for Theoretical Physics, 
Trieste, Italy.}
\end{center}
\vspace{1cm}
\centerline{\bf Abstract}
\baselineskip=18pt
\bigskip

A generalization of the classical Leray-Schauder fixed point theorem, based
on the infinite-dimensional Borsuk-Ulam type antipode construction, is
proposed. Two completely different proofs based on the projection operator
approach and on a weak version of the well known Krein-Milman theorem are
presented.
\vfill
\begin{center}
MIRAMARE -- TRIESTE\\
May 2007\\
\end{center}
\vfill

\newpage

\section{Introduction}

The classical Leray-Schauder fixed point theorem and its diverse versions 
\cite{Go2,AE,DG,Go1,Go3,KF,Ze,Ni} in infinite-dimensional both Banach and
Frechet spaces, being nontrivial generalizations of the well known
finite-dimensional Brouwer fixed point theorem, have many very important
applications \cite{Go2,DG,Go1,Go3,IT,Ge2} in modern applied analysis. In
particular, there exist many problems in theories of differential and
operator equations \cite{Go2,IT,Ru,Ze,Ge2,Ni}, which can be uniformly
formulated as 
\begin{equation}
\hat{a}\text{ }x=f(x),  \label{1.1}
\end{equation}%
where $\hat{a}:E_{1}\rightarrow E_{2}$ is some closed surjective linear
operator from Banach space $E_{1}$ into Banach space $E_{2},$ defined on a
domain $D(\hat{a})\subset E_{1},$\ and $\ f:E_{1}\rightarrow E_{2}$ is some,
in general, nonlinear continuous mapping, whose domain $D(f)\subseteq D(\hat{%
a})\cap S_{r}(0),$ with $S_{r}(0)\subset E_{1}$\ being the sphere of radius $%
r\in \mathbb{R}_{+}$ centered at zero. Concerning the mapping $%
f:E_{1}\rightarrow E_{2}$ we will assume that it is $\hat{a}$\ -compact.
This means that the induced mapping $f_{gr}:D_{gr}(\hat{a})\rightarrow
E_{2}, $ where $D_{gr}(\hat{a})\subset E_{1}\oplus E_{2}$ is the extended
graph domain endowed with the graph-norm, Lipschitz-projected onto the space 
$E_{1} $\ via $j:$\ $D_{gr}(\hat{a})\subset E_{1},$ and the following
equality $\ \ \ f_{gr}(\bar{x})=f(j(\bar{x}))$\ holds for any $\bar{x}\in
D_{gr}(\hat{a}).$ It is easy to observe also \cite{Ge1} that the mapping $%
f:E_{1}\rightarrow E_{2}$ is $\hat{a}$-compact if and only if it is
continuous and for any bounded set $A_{2}\subset E_{2}$ and arbitrary
bounded set $A_{1}\subset D(f) $ the set $f(A_{1}\cap \hat{a}^{-1}(A_{2}))$
is relatively compact in $E_{2}. $ The empty set $\varnothing ,$ by
definition, is considered to be compact too.

\section{\ Preliminary constructions}

Assume that a continuous mapping $f:E_{1}\rightarrow E_{2}$ satisfies the
following conditions:

$1)$\ \ the domain $D(f)=D(\hat{a})\cap S_{r}(0);$

$2)$\ \ the mapping $f:D(f)\rightarrow E_{2}$ is $\hat{a}$ - compact;

$3)$\ there holds a bounded constant $k_{f}>0,$\ such that $\underset{y\in
S_{r}(0)}{\sup }\frac{1}{r}\left\Vert f(y)\right\Vert _{2}=k_{f}^{-1},$ 
\newline
where a linear operator $\hat{a}:E_{1}\rightarrow E_{2}$ is taken closed and
surjective with the domain $D(\hat{a})\subset E_{1}$. The domain $D(\hat{a}%
), $ \ in general, \ can not be dense in $E_{1.}$

Let now $\tilde{E}_{1}:=E_{1}/Ker$ $\hat{a}$ and \ $p_{1}:E_{1}\rightarrow 
\tilde{E}_{1}$ be the corresponding projection. The induced mapping $\tilde{a%
}:\tilde{E}_{1}\rightarrow E_{2}$ with the domain $D(\tilde{a}):=p_{1}(D(%
\hat{a}))$ is defined as usual, that is for any $\tilde{x}\in D(\tilde{a}),$ 
$\hat{a}(\tilde{x}):=a(p_{1}(\tilde{x})).$ \ It is a well know fact \cite%
{AE,KF,Ze} that the mapping $\tilde{a}:\tilde{E}_{1}\rightarrow E_{2}$ is
invertible and its norm is calculated as 
\begin{equation}
\left\Vert \tilde{a}^{-1}\right\Vert :=\underset{\left\Vert y\right\Vert
_{2}=1}{\sup }\left\Vert \tilde{a}^{-1}(y)\right\Vert =\underset{\left\Vert
y\right\Vert _{2}=1}{\sup }\underset{x\in D(\hat{a})}{\inf }\left\{
\left\Vert x\right\Vert _{1}:a(x)=y\right\} ,  \label{2.1a}
\end{equation}%
where we denoted by \ $\left\Vert \cdot \right\Vert _{1}$ and \ $\left\Vert
\cdot \right\Vert _{2}$ the corresponding norms in spaces $E_{1}$ and $%
E_{2}. $ The following standard lemma \cite{KF,Ze} holds.

\begin{lemma}
\label{Lm_2.1}\ The mapping $\tilde{a}:\tilde{E}_{1}\rightarrow E_{2}$ is
invertible and the norm $\left\Vert \tilde{a}^{-1}\right\Vert :=k(\hat{a}%
)<\infty .$
\end{lemma}

\begin{proof}
We have, by definition (\ref{2.1a}), that the norm $\left\Vert \tilde{a}%
^{-1}\right\Vert $ equals 
\begin{equation}
k(\hat{a})=\left\Vert \tilde{a}^{-1}\right\Vert :=\underset{y\in E_{2}}{\sup 
}\frac{\left\Vert \tilde{a}^{-1}(y)\right\Vert _{\tilde{E}_{1}}}{\left\Vert
y\right\Vert _{2}}=\underset{}{\underset{y\in E_{2}}{\sup }\frac{1}{%
\left\Vert y\right\Vert _{2}}}\underset{x\in D(\hat{a})}{\inf }\left\{
\left\Vert x\right\Vert _{1}:\hat{a}(x)=y\right\} .  \label{2.1}
\end{equation}%
Since the linear mapping $\hat{a}:E_{1}\rightarrow E_{2}$ is surjective, the
mapping $\hat{a}^{-1}:E_{2}\rightarrow \tilde{E}_{1}$ is defined on the
whole space $E_{2}.$\ Moreover, as the mapping $\hat{a}:E_{1}\rightarrow
E_{2}$ is a closed operator, the induced inverse operator $\tilde{a}%
^{-1}:E_{2}\rightarrow \tilde{E}_{1}$ is closed \cite{KF,Ru,Ze} too.
Thereby, making use of the classical closed graph theorem \cite{AE,IT,KF},
we conclude that the inverse operator $\tilde{a}^{-1}:E_{2}\rightarrow 
\tilde{E}_{1}$ is bounded, that is norm 
\begin{equation}
\left\Vert \tilde{a}^{-1}\right\Vert :=k(\hat{a})<\infty ,  \label{2.2}
\end{equation}%
finishing the proof.
\end{proof}

The next lemma characterizes the multi-valued mapping $\hat{a}%
^{-1}:E_{2}\rightarrow E_{1}$ by means of the constant $k(\hat{a})<\infty ,$
\ defined by \ (\ref{2.2}).

\begin{lemma}
\label{Lm_2.2}The multi-valued inverse mapping $\hat{a}:E_{2}\rightarrow
E_{1}$ is Lipschitzian with the Lipschitz constant $k(\hat{a})<\infty ,$
that is 
\begin{equation}
\rho _{\chi }(\hat{a}^{-1}(y_{1}),\hat{a}^{-1}(y_{2}))\leq k(\hat{a}%
)\left\Vert y_{1}-y_{2}\right\Vert _{2}  \label{2.3}
\end{equation}%
for any $y_{1},y_{2}\in E_{2},$ where $\rho _{\chi }:\tilde{E}_{1}\times 
\tilde{E}_{1}\rightarrow \mathbb{R}_{+}$ is the standard Hausdorf metrics\ 
\cite{AE,KF,Ze}\ in the space $E_{1}.$
\end{lemma}

\begin{proof}
The statement is a simple corollary from formula \ (\ref{2.1}) and the
Hausdorf metrics definition.
\end{proof}

To describe the solution set of equation \ (\ref{1.1}) we need to know a
more deeper structure of the mapping $\hat{a}:E_{1}\rightarrow E_{2}$ and
its multi-valued inverse $\hat{a}^{-1}:E_{2}\rightarrow E_{1}.$ Namely, we
are interested in finding a suitable, in general, nonlinear continuous
selection $s:E_{2}\rightarrow E_{1}$ \ \cite{AE,IT,Ni,Mi}\ of the
multi-valued mapping $\hat{a}^{-1}:E_{2}\rightarrow E_{1},$ satisfying some
additional properties.

The following theorem is crucial for proving the main result obtained below.

\begin{lemma}
\label{Th_2.3}For any constant $k_{s}>k(\hat{a})$ there exists a continuous
odd mapping \ $s:E_{2}\rightarrow E_{1},$ satisfying the following
conditions: i) $\hat{a}(s(y))=y$\ \ for any $y\in E_{2}$ $;$\ \ ii) $%
\left\Vert s(y)\right\Vert _{1}\leq k_{s}\left\Vert y\right\Vert _{2},$ $%
y\in E_{2}$ $.$
\end{lemma}

\begin{proof}
Since$_{\text{ }}$ the \ multi-valued mapping \ $\hat{a}^{-1}:E_{2}%
\rightarrow E_{1}$ is defined on the whole Banach space $E_{2},$ one can
write down that 
\begin{equation}
\hat{a}^{-1}\text{ }y=\bar{x}_{y}\oplus Ker\text{ }\hat{a}  \label{2.4}
\end{equation}%
for any $y\in E_{2}$ and some specified elements $\bar{x}_{y}\in
E_{1}\backslash Ker$ $\hat{a},$ labelled \ by elements $y\in E_{2}.$ If the
composition \ (\ref{2.4}) \ is already specified, we can define a selection $%
s:$\ $E_{2}\rightarrow E_{1}$ as follows:%
\begin{equation}
s(y):=\frac{1}{2}(\bar{x}_{y}-\bar{x}_{-y})\oplus \frac{1}{2}(\bar{c}_{y}-%
\bar{c}_{-y}),  \label{2.5}
\end{equation}%
where the elements $\bar{c}_{y}\in Ker$ $\hat{a},$ $y\in E_{2},$\ are chosen
arbitrary, but fixed. It is now easy to check that 
\begin{equation}
s(-y)=-s(y)  \label{2.6}
\end{equation}%
and%
\begin{eqnarray}
\hat{a}\text{ }s(y) &=&\hat{a}\text{ }(\frac{1}{2}(\bar{x}_{y}-\bar{x}%
_{-y})\oplus \frac{1}{2}(\bar{c}_{y}-\bar{c}_{-y}))  \label{2.7} \\
&=&\frac{1}{2}\hat{a}\text{ }\bar{x}_{y}-\frac{1}{2}\hat{a}\ \bar{x}_{-y}=%
\frac{1}{2}y-\frac{1}{2}(-y)=y  \notag
\end{eqnarray}%
for all $y\in E_{2},$ thereby the mapping \ \ (\ref{2.5}) \ satisfies the
main conditions $i)$\ and $ii)$\ above. To state the continuity of the
mapping (\ref{2.5}), we will consider below expression \ (\ref{2.1}) for the
norm $\left\Vert \tilde{a}^{-1}\right\Vert =k(\hat{a})$ of the linear
mapping $\tilde{a}^{-1}:E_{2}\rightarrow \tilde{E}_{1}.$ We can easily write
down the following inequality 
\begin{eqnarray}
\left\Vert s(y)\right\Vert _{1} &=&\left\Vert \frac{1}{2}(\bar{x}_{y}-\bar{x}%
_{-y})\oplus \frac{1}{2}(\bar{c}_{y}-\bar{c}_{-y})\right\Vert _{1}
\label{2.8} \\
&=&\frac{1}{2}\left\Vert (\bar{x}_{y}\oplus \bar{c}_{y})-(\bar{x}_{-y}\oplus 
\bar{c}_{-y})\right\Vert _{1}  \notag \\
&\leq &\frac{1}{2}(\left\Vert (\bar{x}_{y}\oplus \bar{c}_{y})\right\Vert
_{1}+\left\Vert (\bar{x}_{-y}\oplus \bar{c}_{-y})\right\Vert _{1})  \notag \\
&\leq &\frac{1}{2}k_{s}\left\Vert y\right\Vert _{2}+\frac{1}{2}%
k_{s}\left\Vert y\right\Vert _{2}=k_{s}\left\Vert y\right\Vert _{2},  \notag
\end{eqnarray}%
giving rise to the continuity of mapping (\ref{2.5}), where we have assumed
that there exists such a constant $k_{s}>0,$ \ that \ 
\begin{equation}
\left\Vert (\bar{x}_{y}\oplus \bar{c}_{y})\right\Vert _{1}\leq
k_{s}\left\Vert y\right\Vert _{2},  \label{2.9}
\end{equation}%
for all $y\in E_{2}.$ This constant $k_{s}>k(\hat{a})$ strongly depends on
the choice of elements \ \ $\bar{c}_{y}\in Ker$ $\hat{a},$ $y\in E_{2},$
what one can observe from definition \ (\ref{2.1}). \ Really, owing to the
definition of infimum, for any $\varepsilon >0$\ and all $y\in E_{2}$\ there
exist elements $\bar{x}_{y}^{(\varepsilon )}\oplus \bar{c}_{y}^{(\varepsilon
)}\in E_{1},$ such that \ 
\begin{equation}
k(\hat{a})\leq \frac{\left\Vert \bar{x}_{y}^{(\varepsilon )}\oplus \bar{c}%
_{y}^{(\varepsilon )}\right\Vert _{1}}{\left\Vert y\right\Vert _{2}}<k(\hat{a%
})+\varepsilon :=k_{s}.  \label{2.10}
\end{equation}

Now making now use of formula \ \ (\ref{2.5}), we can construct a selection $%
s_{\varepsilon }:E_{2}\rightarrow E_{1}$ as follows:%
\begin{equation}
s_{\varepsilon }(y):=\frac{1}{2}(\bar{x}_{y}^{(\varepsilon )}-\bar{x}%
_{-y}^{(\varepsilon )})\oplus \frac{1}{2}(\bar{c}_{y}^{(\varepsilon )}-\bar{c%
}_{-y}^{(\varepsilon )}),  \label{2.11}
\end{equation}%
satisfying, owing to inequalities \ (\ref{2.10}), \ the searched for
conditions $i)$\ and $ii)$:%
\begin{equation}
\hat{a}\text{ }s_{\varepsilon }(y)=y,\text{ \ \ \ \ \ \ \ }\left\Vert
s_{\varepsilon }(y)\right\Vert _{1}\leq k_{s}\left\Vert y\right\Vert _{2}
\label{2.12}
\end{equation}%
for all $y\in E_{2}$ and $k_{s}:=k(\hat{a})+\varepsilon $ $,$ $\varepsilon
>0.$\newline
Moreover, the mapping $s_{\varepsilon }:E_{2}\rightarrow E_{1}$ is, by
construction, continuous \cite{Mi,Dz,Ge1} and odd that finishes the proof. \ 
\end{proof}

\section{ An infinite - dimensional Borsuk-Ulam type generalization of the
Leray-Schauder fixed point theorem}

Consider now the equation \ (\ref{1.1}), where mappings \ $\hat{a}$ $%
:E_{1}\rightarrow E_{2}$ \ and $f:E_{1}\rightarrow E_{2}$ satisfy the
conditions described above. Moreover, we will assume that the selection $%
s:E_{2}\rightarrow E_{1},$ constructed above, and the mapping $f:D(f)\subset
E_{1}\rightarrow E_{2}$ \ satisfy additionally the following inequalities:%
\begin{equation}
k(\hat{a})<k_{s}<k_{f}\text{ },  \label{3.1a}
\end{equation}%
where, by definition, 
\begin{equation}
\underset{x\in S_{r}(0)}{\sup }\frac{1}{r}\left\Vert f(x)\right\Vert
:=k_{f}^{-1}<\infty .  \label{3.1b}
\end{equation}%
Then the following main theorem holds.

\begin{theorem}
\label{Th_3.1}Assume that the dimension $\dim Ker$ $\hat{a}\geq 1,$ then
equation (\ref{1.1})\ possesses on the sphere $S_{r}(0)\subset E_{1}$ the
nonempty solution set $\mathcal{N}(\hat{a},f)\subset E_{1},$ whose
topological dimension $\dim \mathcal{N}(\hat{a},f)\geq \dim Ker$ $\hat{a}-1.$
\end{theorem}

\begin{proof}
Suppose that $\dim Ker$\ $\hat{a}\geq 1$\ \ and state first that the set $%
\mathcal{N}(\hat{a},f)$\ is nonempty. Consider a reduced mapping $f_{r}:D(%
\hat{a})\subset E_{1}\rightarrow E_{2},$ where 
\begin{equation}
f_{r}(x):=\left\{ 
\begin{array}{c}
\frac{\left\Vert x\right\Vert _{1}}{r}f(\frac{rx}{\left\Vert x\right\Vert
_{1}}),\text{ if \ }x\neq 0 \\ 
0,\text{ \ \ \ \ \ \ \ \ \ \ \ \ \ \ \ if \ }=0%
\end{array}%
\right\}  \label{3.1}
\end{equation}%
and observe that this mapping is \ $\hat{a}$ - compact too, if the mapping $%
f:D(f)\subset E_{1}\rightarrow E_{2}$ was taken $\hat{a}$\ - compact.
Really, for any bounded sets $A_{2}\subset E_{2}$ \ and $A_{1}\subset
B_{R}(0)\cap D(\hat{a})$ the set 
\begin{equation}
f_{r}(A_{1}\cap \hat{a}^{-1}(A_{2}))\subset \left\{ ty\in E_{2}:t\in \left[
0,R/r\right] ,y\in f(S_{r}(0))\cap \hat{a}^{-1}(A_{2})\right\} :=F_{r}
\label{3.2}
\end{equation}%
is relatively compact owing to the $\hat{a}$ - compactness of the mapping $%
f:D(f)\subset E_{1}\rightarrow E_{2},$ where \ $B_{R}(0)$ is a ball of
radius $R>0.$ Thereby, the closed set $\bar{F}_{r}\subset E_{2}$\ is
compact, or the mapping \ (\ref{3.1}) is $\hat{a}$ - compact.

Assume now that a mapping $s:E_{2}\rightarrow E_{1}$ satisfies all of the
conditions formulated in Theorem \ref{Th_2.3}. Take a nonzero element $\bar{c%
}\in Ker$\ $\hat{a},$ define the Banach space \ $E_{2}^{(+)}:=E_{2}\oplus 
\mathbb{R}$\ \ and consider a set of mappings $\varphi _{r}^{(\varepsilon
)}: $\ $E_{2}^{(+)}\rightarrow E_{2},$\ \ where 
\begin{equation}
\varphi _{r}^{(\varepsilon )}(y,t):=\frac{t}{t^{2}+\varepsilon ^{2}}%
f_{r}(ts(y)+t^{2}\bar{c})  \label{3.3}
\end{equation}%
for all $(y,t)\in E_{2}^{(+)},$\ small enough $\varepsilon \in \mathbb{R}%
\backslash \left\{ 0\right\} $ and some fixed nontrivial element $\bar{c}\in
Ker$\ $\hat{a}.$\ It is also evident that 
\begin{equation}
\varphi _{r}^{(\varepsilon )}(y,0):=0,  \label{3.4}
\end{equation}%
being well definite for all $\varepsilon \in \mathbb{R}\backslash \left\{
0\right\} $ and $y\in E_{2},$\ owing to condition $3)$\ imposed above on the
mapping $f:D(f)\subset E_{1}\rightarrow E_{2}.$ The set of mappings \ (\ref%
{3.3})\ is, evidently, odd, that is 
\begin{equation}
-\varphi _{r}^{(\varepsilon )}(y,t)=\varphi _{r}^{(\varepsilon )}(-y,-t)
\label{3.5}
\end{equation}%
for all $(y,t)\in E_{2}^{(+)},$ \ $\varepsilon \in \mathbb{R}\backslash
\left\{ 0\right\} $ and moreover, it is compact. Really, for any bounded set 
$A_{2}^{(+)}:=A_{2}\oplus \Delta \subset E_{2}^{(+)},$\ where \ $\Delta
\subset \mathbb{R}$ is an arbitrary bounded interval, the set $B_{2}:=%
\underset{t\in \Delta }{\cup }B_{2}^{(t)},$ $\ B_{2}^{(t)}:=\left\{ s(y)+t%
\bar{c}\in E_{2}\right\} ,$ is \ bounded too, and $B_{2}\subset \hat{a}%
^{-1}(A_{2}).$ Owing to the $\hat{a}$ - compactness of mapping \ (\ref{3.1}%
), one gets that the set \ 
\begin{equation}
\varphi _{r}^{(\varepsilon )}(A_{2}^{(+)})=\underset{t\in \Delta }{\cup }%
\frac{t}{t^{2}+\varepsilon ^{2}}f_{r}(tB_{2}^{(t)})  \label{3.6}
\end{equation}%
is relatively compact, since all of the sets $f_{r}(tB_{2}^{(t)})\subset
E_{2}\ $are relatively compact for any $t\in \Delta $ \ and, owing to the
condition $3)$\ mentioned above, the set $\varphi _{r}^{(\varepsilon
)}(A_{2}^{(+)})$\ is bounded for any $\varepsilon \in \mathbb{R}\backslash
\left\{ 0\right\} .$ Thereby, \ the closed set $\overline{\varphi
_{r}^{(\varepsilon )}(A_{2}^{(+)})}\subset E_{2}$ for any $\varepsilon \in 
\mathbb{R}\backslash \left\{ 0\right\} ,$ meaning that the mapping \ (\ref%
{3.3}) is compact.

Take now the unit sphere $S_{1}^{(+)}(0)\subset E_{2}^{(+)}$\ and consider
the equation 
\begin{equation}
\varphi _{r}^{(\varepsilon )}(y,t)=y  \label{3.7}
\end{equation}%
for $(y,t)\in S_{1}^{(+)}(0)$ \ and $\varepsilon \in \mathbb{R}\backslash
\left\{ 0\right\} $ that is 
\begin{equation}
\left\Vert y\right\Vert _{2}^{2}+t^{2}=1.  \label{3.8}
\end{equation}

We assert that equation \ (\ref{3.7}) \ possesses for any $\varepsilon \in 
\mathbb{R}\backslash \left\{ 0\right\} $ a solution $(y_{\varepsilon
},t_{\varepsilon })\in S_{1}^{(+)}(0),$ such that $t_{\varepsilon }\neq 0$ \
and 
\begin{equation}
\frac{t_{\varepsilon }}{t_{\varepsilon }^{2}+\varepsilon ^{2}}%
f_{r}(t_{\varepsilon }s(y_{\varepsilon })+t_{\varepsilon }^{2}\bar{c}%
)=y_{\varepsilon }\text{ },  \label{3.9}
\end{equation}%
where the vector \ $t_{\varepsilon }s(y_{\varepsilon })+t_{\varepsilon }^{2}%
\bar{c}\in E_{2}$ \ is nontrivial (i.e. it is not equal to zero!). This is
guaranteed by conditions imposed on the mapping $f:S_{r}(0)\subset
E_{1}\rightarrow E_{2}$ and the following Borsuk-Ulam type theorem,
generalizing the well known Borsuk-Ulam \cite{AE,Ni,Ze,Go1} antipode
theorem, proved in \cite{Ge1} \ and \ formulated below in a convenient for
us form.

\begin{theorem}
\label{Th_3.2}Let $E_{2}^{(+)}$ and $E_{2}$ be Banach spaces, $\hat{b}%
:E_{2}^{(+)}\rightarrow E_{2}$ be a linear continuous surjective operator, $%
S_{r}^{(+)}(0)\subset E_{2}^{(+)}$ \ be a sphere of radius $r>0$ centered at
zero of $E_{2}^{(+)}$ and $\varphi :S_{r}^{(+)}(0)\rightarrow E_{2}$ be a
compact, in general nonlinear, odd mapping. Then if $\dim Ker$ $\hat{b}\geq
1,$ the equation 
\begin{equation}
\hat{b}\text{ }z=\varphi (z),  \label{3.9a}
\end{equation}%
$z\in S_{r}^{(+)}(0),$ possesses the nonempty solution set $\mathcal{N}(\hat{%
b},\varphi )\subset $ $E_{2}^{(+)},$ whose topological dimension $\dim 
\mathcal{N}(\hat{b},\varphi )\geq \dim Ker$ $\hat{b}-1.$
\end{theorem}
\end{proof}

\begin{proof}
To state that our equation \ (\ref{3.7}) is solvable, it is enough to define
a suitable linear, bounded and surjective operator $\hat{b}%
:E_{2}^{(+)}\rightarrow E_{2}$ and apply Theorem \ref{Th_3.2}. Put, by
definition, 
\begin{equation}
\hat{b}\text{ }z:=y,  \label{3.9b}
\end{equation}%
where $z:=(y,t)\in E_{2}^{(+)},$ $y\in E_{2},$ $t\in \mathbb{R}.$ The
operator \ (\ref{3.9b}) is evidently linear bounded with the norm $||\hat{b}%
||=1$ and surjective with $Range$\ $\hat{b}=E_{2}.$ Take now the mapping $%
\varphi :=\varphi _{r}^{(\varepsilon )}:E_{2}^{(+)}\rightarrow E_{2}$ for $%
\varepsilon \in \mathbb{R}\backslash \left\{ 0\right\} $ and apply Theorem %
\ref{Th_3.1}. Since $\dim Ker$ $\hat{b}=1,$ we get that equation \ (\ref{3.7}%
), written in the form 
\begin{equation}
\varphi (z):=\varphi _{r}^{(\varepsilon )}(z)=\hat{b}\text{ }z  \label{3.9c}
\end{equation}%
for all $z\in E_{2}^{(+)},$ possesses a nonempty solution set $\mathcal{N}(%
\hat{b},\varphi _{r}^{(\varepsilon )})\subset $ $E_{2}^{(+)},$\ whose
topological dimension \ $\dim \mathcal{N}(\hat{b},\varphi _{r}^{(\varepsilon
)})\geq 0$ \ for all $\varepsilon \in \mathbb{R}\backslash \left\{ 0\right\}
.$ Assume now, for a moment, that the value $t_{\varepsilon }\neq 0.$ Then,
based on expression \ \ (\ref{3.9}), one can easily get that the
well-defined vector 
\begin{equation}
x_{\varepsilon }:=\frac{rt_{\varepsilon }(s(y_{\varepsilon })+t_{\varepsilon
}\bar{c})}{|t_{\varepsilon }|\left\Vert s(y_{\varepsilon })+t_{\varepsilon }%
\bar{c}\right\Vert _{1}}\text{ }  \label{3.10}
\end{equation}%
satisfies the following equation:%
\begin{equation}
f(x_{\varepsilon })=t_{\varepsilon }^{-2}(t_{\varepsilon }^{2}+\varepsilon
^{2})\hat{a}\text{ }x_{\varepsilon }.  \label{3.11}
\end{equation}%
Really, from (\ref{3.9}) we obtain that%
\begin{eqnarray}
\frac{t_{\varepsilon }}{t_{\varepsilon }^{2}+\varepsilon ^{2}}%
f_{r}(t_{\varepsilon }s(y_{\varepsilon })+t_{\varepsilon }^{2}\bar{c}) &=&%
\frac{t_{\varepsilon }\left\vert t_{\varepsilon }\right\vert \left\Vert
s(y_{\varepsilon })+t_{\varepsilon }\bar{c}\right\Vert _{1}}{%
r(t_{\varepsilon }^{2}+\varepsilon ^{2})}\ f\left( \frac{rt_{\varepsilon
}(s(y_{\varepsilon })+t_{\varepsilon }\bar{c})}{|t_{\varepsilon }|\left\Vert
s(y_{\varepsilon })+t_{\varepsilon }\bar{c}\right\Vert _{1}}\right)  \notag
\\
&=&\frac{t_{\varepsilon }\left\vert t_{\varepsilon }\right\vert \left\Vert
s(y_{\varepsilon })+t_{\varepsilon }\bar{c}\right\Vert _{1}}{%
r(t_{\varepsilon }^{2}+\varepsilon ^{2})}f(x_{\varepsilon })=y_{\varepsilon
}.  \label{3.12}
\end{eqnarray}%
Whence, recalling the identity \ $\hat{a}(s(y_{\varepsilon
}))=y_{\varepsilon }$ \ for any \ $y_{\varepsilon }\in E_{2},$ we find that 
\begin{eqnarray}
f(x_{\varepsilon }) &=&\frac{(t_{\varepsilon }^{2}+\varepsilon ^{2})r\text{ }%
\hat{a}\text{ }(s(y_{\varepsilon }))}{t_{\varepsilon }\left\Vert
s(y_{\varepsilon })+t_{\varepsilon }\bar{c}\right\Vert _{1}}=\frac{%
(t_{\varepsilon }^{2}+\varepsilon ^{2})}{t_{\varepsilon }^{2}}\text{ }\hat{a}%
\text{ }\left( \frac{rs(y_{\varepsilon })t_{\varepsilon }}{|t_{\varepsilon
}|\left\Vert s(y_{\varepsilon })+t_{\varepsilon }\bar{c}\right\Vert _{1}}%
\right)  \notag \\
&=&\frac{(t_{\varepsilon }^{2}+\varepsilon ^{2})}{t_{\varepsilon }^{2}}\text{
}\hat{a}\text{ }\left( \frac{t_{\varepsilon }r(s(y_{\varepsilon
})+t_{\varepsilon }\bar{c})\ }{|t_{\varepsilon }|\left\Vert s(y_{\varepsilon
})+t_{\varepsilon }\bar{c}\right\Vert _{1}}\right) =\frac{(t_{\varepsilon
}^{2}+\varepsilon ^{2})}{t_{\varepsilon }^{2}}\text{ }\hat{a}\text{ }%
x_{\varepsilon },  \label{3.13}
\end{eqnarray}%
where we took into account the linearity of the operator $\hat{a}%
:E_{1}\rightarrow E_{2}$ \ and \ the fact that the vector $\bar{c}\in Ker$ $%
\hat{a}.$ Thereby, \ the constructed vector \ $x_{\varepsilon }\in E_{1}$
satisfies for $\varepsilon \in \mathbb{R}\backslash \left\{ 0\right\} $ the
equation (\ref{3.11}).\ The considerations above hold since we assumed that $%
t_{\varepsilon }\neq 0$ for all $\varepsilon \in \mathbb{R}\backslash
\left\{ 0\right\} .$ To show this is the case, assume the inverse that is $%
t_{\varepsilon }=0$ for some $\varepsilon \in \mathbb{R}\backslash \left\{
0\right\} .$ We then get from (\ref{3.9}) \ and condition $2)$\ imposed
before on the mapping $f:D(f)\subset E_{1}\rightarrow E_{2}$ right away that
simultaneously there should be fulfilled the equality $\left\Vert
y_{\varepsilon }\right\Vert _{2}=0,$\ contradicting to the condition \ (\ref%
{3.8}). \ Thus, for all $\varepsilon \in \mathbb{R}\backslash \left\{
0\right\} $ \ the value $t_{\varepsilon }\neq 0.$\ If to state more accurate
estimations, mainly, that the following inequalities%
\begin{equation}
1>\underset{\varepsilon \rightarrow 0}{\underline{\lim }}\left\vert
t_{\varepsilon }\right\vert ^{2}\geq 1-\alpha _{0}^{2}>0  \label{3.14}
\end{equation}%
hold for some positive value $\alpha _{0}>0,$ then one can try to calculate
the limit:%
\begin{equation}
\underset{n{\rightarrow }\infty }{\lim }f(x_{\varepsilon _{n}})=f(x_{0})=%
\underset{n\rightarrow \infty }{\lim }\left( t_{\varepsilon
_{n}}^{-2}(t_{\varepsilon _{n}}^{2}+\varepsilon _{n}^{2})\text{ }\hat{a}%
\text{ }x_{\varepsilon _{n}}\right) =\hat{a}\text{ }x_{0}\text{ }
\label{3.15}
\end{equation}%
for some subsequence \ $\varepsilon _{n}\rightarrow 0$ \ as $n\rightarrow
\infty $. \ Here we have assumed that there exists $\underset{{n}\rightarrow
\infty }{\lim }x_{\varepsilon _{n}}=x_{0},$ \ that is 
\begin{equation}
\underset{n\rightarrow \infty }{\lim }\frac{t_{\varepsilon
_{n}}r(s(y_{\varepsilon _{n}})+t_{\varepsilon _{n}}\bar{c})\ }{\left\vert
t_{\varepsilon _{n}}\right\vert \left\Vert s(y_{\varepsilon
_{n}})+t_{\varepsilon _{n}}\bar{c}\right\Vert _{1}}=x_{0}  \label{3.16}
\end{equation}%
depending on the chosen before nontrivial vector $\bar{c}\in Ker$ $\hat{a}.$

Owing to the $\hat{a}$-compactness of the mapping $f:D(f)\subset
E_{1}\rightarrow E_{2}$ and the continuity of the operators $\tilde{a}%
^{-1}:E_{2}\rightarrow \tilde{E}_{1}$ \ and $s:E_{2}\rightarrow E_{1},$ for
the limit \ (\ref{3.16}) \ to exist it is enough only to \ state \ that \
there holds inequality \ (\ref{3.14}). Really, since owing to relationship (%
\ref{3.8}) for all $\varepsilon >0$ the following condition 
\begin{equation}
\left\vert t_{\varepsilon }\right\vert ^{2}+\left\Vert y_{\varepsilon
}\right\Vert _{2}^{2}=1  \label{3.17}
\end{equation}%
holds, the limit (\ref{3.16}) \ will exist, if to state equivalently that 
\begin{equation}
\underset{{n}\rightarrow \infty }{\lim }\left\Vert y_{\varepsilon
_{n}}\right\Vert _{2}\leq \alpha _{0}<1.  \label{3.18}
\end{equation}%
To show inequality \ \ (\ref{3.18}), consider expression \ (\ref{3.9}) \ and
make the following estimations: 
\begin{eqnarray}
\underset{n\rightarrow \infty }{\lim }\left\Vert y_{\varepsilon
_{n}}\right\Vert _{2} &=&\underset{n\rightarrow \infty }{\lim }\left( \frac{%
\left\vert t_{\varepsilon _{n}}\right\vert }{t_{\varepsilon
_{n}}^{2}+\varepsilon _{n}^{2}}\left\Vert f_{r}(t_{\varepsilon
_{n}}s(y_{\varepsilon _{n}})+t_{\varepsilon _{n}}^{2}\bar{c})\right\Vert
_{2}\right)  \notag \\
&\leq &\underset{n\rightarrow \infty }{\lim }\left( \frac{\left\vert
t_{\varepsilon _{n}}\right\vert ^{2}}{(t_{\varepsilon _{n}}^{2}+\varepsilon
_{n}^{2})}\frac{\left\Vert s(y_{\varepsilon _{n}})+t_{\varepsilon _{n}}\bar{c%
}\right\Vert _{1}}{r}f\left( \frac{\ rt_{\varepsilon _{n}}(s(y_{\varepsilon
_{n}})+t_{\varepsilon _{n}}\bar{c})\ }{\left\vert t_{\varepsilon
_{n}}\right\vert \left\Vert s(y_{\varepsilon _{n}})+t_{\varepsilon _{n}}\bar{%
c}\right\Vert _{1}}\right) \right)  \notag \\
&\leq &\underset{n\rightarrow \infty }{\lim }\left\Vert s(y_{\varepsilon
_{n}})+t_{\varepsilon _{n}}\bar{c}\right\Vert _{1}k_{f}^{-1}\leq k_{f}^{-1}(%
\underset{n\rightarrow \infty }{\lim }\left\Vert s(y_{\varepsilon
_{n}})\right\Vert _{1}+(1-\underset{n\rightarrow \infty }{\lim }\left\Vert
y_{\varepsilon _{n}}\right\Vert _{2}^{2})^{1/2}\left\Vert \bar{c}\right\Vert
_{1})  \label{3.19} \\
&\leq &k_{f}^{-1}(k_{s}\underset{n\rightarrow \infty }{\lim }\left\Vert
y_{\varepsilon _{n}}\right\Vert _{2}+[1-\underset{n\rightarrow \infty }{\lim 
}\left\Vert y_{\varepsilon _{n}}\right\Vert _{2}^{2}]^{1/2}\left\Vert \bar{c}%
\right\Vert _{1}).  \notag
\end{eqnarray}%
Thus, we obtain from \ (\ref{3.19}) \ that the value $\ \ \ \alpha _{0}:=%
\underset{n\rightarrow \infty }{\lim }\left\Vert y_{\varepsilon
_{n}}\right\Vert _{2}\in \mathbb{R}_{+}$ \ satisfies the following
inequalities:%
\begin{equation}
0\leq \alpha _{0}\leq k_{f}^{-1}(k_{s}\text{ }\alpha _{0}\text{ }+\text{ }%
(1-\alpha _{0}^{2})^{1/2}\left\Vert \bar{c}\right\Vert _{1})\leq 1
\label{3.20}
\end{equation}%
where, in general, $\alpha _{0}\in \lbrack 0,1]$ $.$\ For inequalities \ (%
\ref{3.20}) \ to hold true, we need to consider two possibilities:%
\begin{equation}
a)\text{ \ }k_{s}k_{f}^{-1}\geq 1\text{ };\text{ \ \ \ \ }b)\text{ }%
k_{s}k_{f}^{-1}<1.  \label{3.21}
\end{equation}%
For the case $a)$\ of \ (\ref{3.21}) we can easily state that 
\begin{equation}
1\leq \min (\frac{k_{s}}{k_{f}},1)\leq \alpha _{0}\leq k_{f}^{-1}\sqrt{%
k_{s}^{2}+\left\Vert \bar{c}\right\Vert _{1}^{2}}\text{ }.  \label{3.21a}
\end{equation}%
For the case $b)$\ of \ (\ref{3.21a}) \ one gets similarly that \ \ \ 
\begin{equation}
0\leq \alpha _{0}\leq \frac{\left\Vert \bar{c}\right\Vert _{1}}{\sqrt{\
\left\Vert \bar{c}\right\Vert _{1}^{2}+(k_{s}-k_{f})^{2}}}\text{ }.
\label{3.22}
\end{equation}%
Since we are interested in any value of $\alpha _{0}<1,$ \ the only
inequality \ (\ref{3.22}) \ fits to the searched for exact inequality 
\begin{equation}
0\leq \alpha _{0}\leq \frac{\left\Vert \bar{c}\right\Vert _{1}}{\sqrt{\
\left\Vert \bar{c}\right\Vert _{1}^{2}+(k_{s}-k_{f})^{2}}}<1 \\
,  \label{3.23}
\end{equation}%
guaranteeing the existence of a nontrivial (not zero!) solution to equation\
(\ref{3.15}). Thereby, the nontrivial vector $x_{0}\in D(f)$ constructed
above satisfies, following from \ (\ref{3.15}), the equality \ \ 
\begin{equation}
f(x_{0})=\hat{a}\text{ }x_{0}.  \label{3.24}
\end{equation}%
Moreover, since the vector $x_{0}\in D(f),$ owing to representation \ (\ref%
{3.16}), \ depends nontrivially on the chosen vector $\ \bar{c}\in Ker$ $%
\hat{a},$ \ we deduce that the corresponding to \ (\ref{3.24}) \ solution
set $\mathcal{N}(\hat{a},f)\subset $ $E_{1}$ is nonempty, if $\dim Ker$ $%
\hat{a}$ $\geq 1,$ and the topological dimension $\dim \mathcal{N}(\hat{a}%
,f)\geq \dim Ker$ $\hat{a}$ $-1.$ The latter finishes the proof of the
theorem.
\end{proof}

\section{Corollaries}

The classical Leray-Schauder fixed point theorem, as is well known \cite%
{AE,Go2,KF,Ni,Ze}, \ reads as follows.

\begin{theorem}
\label{Th_4.1}Let a compact mapping $\bar{f}:B\rightarrow B$ in a Banach
space $B$\ is such that there exists a closed convex and bounded set $%
M\subset B,$ for which $\bar{f}(M)\subseteq M.$ Then there exists a fixed
point $\bar{x}\in M,$ such that 
\begin{equation}
\bar{f}(\bar{x})=\bar{x}.\text{ }  \label{4.1}
\end{equation}
\end{theorem}

\begin{proof}
One can present two completely different approaches to the proof of this
classical Leray-Schauder theorem, using the main Theorem \ref{Th_3.1}. The
first one is based on simple geometrical considerations, and the second one,
requires some topological backgrounds.
\end{proof}

\begin{proof}
\textbf{Approach 1.} Put, by definition, that $E_{1}:=B\oplus \mathbb{R},$ \
\ $E_{2}:=B$ and $M_{f}:=Conv$ $\bar{f}(M)\subseteq M$ \ is the convex and
compact convex hull of the image $\bar{f}(M)\subseteq M.$ For any point $%
x\in B$\ one can define the set-valued projection mapping 
\begin{equation}
B\ni x\rightarrow P_{M_{f}}(x)\subset M_{f}\subset B,  \label{4.2}
\end{equation}%
where 
\begin{equation}
\underset{y\in M_{f}}{\inf }||x-y||:=||x-P_{M_{f}}(x)||.  \label{4.3}
\end{equation}%
The set-valued mapping (\ref{4.2}) is well defined and upper semi-continuous 
\cite{BMW,BD} owing to the closedness, boundedness and convexity of the set $%
M_{f}\subset B.$\ Now take the unit sphere $S_{1}(0)$ $\subset E_{1}$ and
construct a mapping $f:S_{1}(0)$ $\subset E_{1}\rightarrow E_{2},$\ where,
by definition, for any $(x,\tau )\in S_{1}(0)$\ 
\begin{equation}
f(x,\tau ):=\bar{f}(\bar{P}_{M_{f}}(x))-\bar{P}_{M_{f}}(x)+\hat{b}\text{ }x,
\label{4.4}
\end{equation}%
$\bar{P}_{M_{f}}:B\rightarrow M_{f}\subset B$ is a suitable continuous
selection \cite{Mi} for the mapping (\ref{4.2}) and $\hat{b}:B\rightarrow B$
\ is an arbitrary compact and surjective mapping. Concerning the
corresponding mapping $\hat{a}:E_{1}\rightarrow E_{2},$ we put, by
definition, 
\begin{equation}
\hat{a}\text{ }(x,\tau ):=\hat{b}\text{ }x  \label{4.5}
\end{equation}%
for all $(x,\tau )\in E_{1}=B\oplus \mathbb{R}.$ \ It is now easy to observe
that the following lemma holds.

\begin{lemma}
\label{Lm_4.1}The mapping $f:S_{1}(0)$ $\subset E_{1}\rightarrow E_{2},$
defined by (\ref{4.4}), is continuous and $\hat{a}-$compact.
\end{lemma}

\begin{proof}
Really, for any $x\in B$\ the element $\ \bar{P}_{M_{f}}(x)\in M_{f}$\ and \ 
$\bar{f}(\bar{P}_{M_{f}}(x))\in M_{f},$\ owing to the invariance $\bar{f}%
(M)\subseteq M.$\ From the compactness \ of the mappings $\bar{f}$\ $:M$\ $%
\rightarrow M$ and $\ \hat{b}:B\rightarrow B$\ one easily gets the $\hat{a}$%
-compactness \ of the constructed mapping $\ f:E_{1}\rightarrow E_{2}$ that
proves the lemma.
\end{proof}

Now taking into account Lemma \ref{Lm_4.1} and the fact that operator $\hat{a%
}:E_{1}\rightarrow E_{2},$\ defined by (\ref{4.5}), is closed and
surjective, owing to the assumptions done above, we can apply to the
equation 
\begin{equation}
\hat{a}\text{ }(x,\tau )=f(x,\tau ),  \label{4.6}
\end{equation}%
where $(x,\tau )\in S_{1}(0)\subset E_{1},$\ the main Theorem \ref{Th_3.1}
and, thereby, state that the corresponding solution set $\mathcal{N}(\hat{a}%
,f)$\ $\subset E_{1}$\ is nonempty, since $\dim Ker$\ $\hat{a}\geq 1.$\ In
particular, from (\ref{4.6}) one gets that 
\begin{equation}
\bar{f}(\bar{P}_{M_{f}}(x_{\tau }))=\bar{P}_{M_{f}}(x_{\tau })  \label{4.7}
\end{equation}%
for the vector $\bar{P}_{M_{f}}(x_{\tau })\in M_{f},$\ where a point $%
x_{\tau }\in B_{1}(0)$\ satisfies the condition $||x_{\tau }||^{2}+||\tau
||^{2}=1$\ for some value $|\tau |\leq 1.$\ 

Thereby, we have stated that the fixed point problem (\ref{4.1}) is solvable
and its solution can, in particular, be obtained as the projection $\bar{x}:=%
\bar{P}_{M_{f}}(x_{\tau })$ of some point $x_{\tau }\in B_{1}(0)$\ upon the
compact, convex and invariant set $M_{f}\subseteq M\subset B.$

\textbf{Approach 2.} We shall start from the following result \cite{RR,Fe}
about the general structure of compact and convex sets in metrizible locally
convex topological vector spaces, being a weak version of the well known
Krein-Milman theorem.

\begin{lemma}
\label{Lm_4.2}Let $E$\ \ be a metrizible locally convex topological vector
space over the fileld $\mathbb{R},$ $\ F\subset E$\ be its dense vector
subspace and $M\subset E$\ be any convex and closed compact subset. Then
there exists a countable linearly independent sequence $\{e_{n}\in F:n\in 
\mathbb{Z}_{+}\},$\ such that $\underset{n\rightarrow \infty }{lim}$\ $%
e_{n}\rightarrow 0,$\ a countable sequence $\{\lambda _{n}(x)\in \mathbb{R}%
:n\in \mathbb{Z}_{+}\},$\ such that 
\begin{equation}
\sum_{n\in \mathbb{Z}_{+}}|\lambda _{n}(x)|\leq 1,  \label{4.8}
\end{equation}%
and every element \ $x\in M$\ allows the representation 
\begin{equation}
x=\sum_{n\in \mathbb{Z}_{+}}\lambda _{n}(x)e_{n}.  \label{4.9}
\end{equation}
\end{lemma}

\begin{proof}
A proof of this lemma can be found, for instance, in \cite{RR,Fe}, so we
will not present it here.
\end{proof}

As any Banach space $B$\ \ is a metrizible locally convex topological vector
space, representation (\ref{4.9}) naturally generates a well-defined
surjective and continuous compact mapping $\xi :l_{1}(\mathbb{Z}_{+};\mathbb{%
R})\rightarrow $ $M_{f}$\ \ $\subset B$ with the domain $D(\xi )=\bar{B}%
_{1}(0),$ where the set $\bar{B}_{1}(0)\subset l_{1}(\mathbb{Z}_{+};\mathbb{R%
})$ is the unit ball centered at zero in the Banach space $l_{1}(\mathbb{Z}%
_{+};\mathbb{R})$ and $\ M_{f}:=Conv$ $\bar{f}(M)\subseteq M$ \ is, as
before, the convex and compact convex hull of the image $\bar{f}(M)\subseteq
M.$ \ The next lemma follows from Lemma \ref{Lm_4.2} and \cite{RR,Fe} and
some related results about the continuous selections from \cite%
{Go1,Go2,IT,Ze}.

\begin{lemma}
\label{Lm_4.3}There exists such a continuous selection $\xi
_{s}^{-1}:B\supset M_{f}\rightarrow \bar{B}_{1}(0)\subset l_{1}(\mathbb{Z}%
_{+};\mathbb{R}),$ $\xi \cdot \xi _{s}^{-1}=id:M_{f}\rightarrow M_{f},$ that
for any vector $x\in M_{f}$ \ the value \ $\xi _{s}^{-1}(x)\in \bar{B}%
_{1}(0) $ determines uniquely this vector by means of representation (\ref%
{4.9}) as 
\begin{equation}
x=\sum_{n\in \mathbb{Z}_{+}}(\xi _{s}^{-1}(x))_{n}e_{n}.  \label{4.10}
\end{equation}%
Moreover, this selection can be chosen in such a way, that an induced
mapping $\bar{F}_{s}:l_{1}(\mathbb{Z}_{+};\mathbb{R})\supset $ $\bar{B}%
_{1}(0)\rightarrow $ $\bar{B}_{1}(0)\subset l_{1}(\mathbb{Z}_{+};\mathbb{R}%
), $ defined as \ 
\begin{equation}
\bar{F}_{s}(\lambda ):=\xi _{s}^{-1}\cdot \bar{f}(\xi (\lambda ))
\label{4.11}
\end{equation}%
for any $\lambda \in \bar{B}_{1}(0)\subset l_{1}(\mathbb{Z}_{+};\mathbb{R}),$
is continuous and \ also compact.
\end{lemma}

\begin{proof}
Modulo the existence \cite{Mi,BMW} of a selection $\xi _{s}^{-1}:B\supset
M_{f}\rightarrow \bar{B}_{1}(0)\subset l_{1}(\mathbb{Z}_{+};\mathbb{R}),$ a
proof is based both on representations (\ref{4.10}) and (\ref{4.11}) and \
on the compactness of the mapping $\xi :l_{1}(\mathbb{Z}_{+};\mathbb{R}%
)\supset \bar{B}_{1}(0)\rightarrow $ $M_{f}$\ \ $\subset B$ and the set $%
M_{f}$ $,$ as well as on the standard fact \cite{KF,Ze} that the continuous
image of a compact set is compact too.
\end{proof}

Pose now the fixed point problem for the compact mapping $\bar{F}_{s}:l_{1}(%
\mathbb{Z}_{+};\mathbb{R})\supset $ $\bar{B}_{1}(0)\rightarrow $ $\bar{B}%
_{1}(0)\subset l_{1}(\mathbb{Z}_{+};\mathbb{R})$ constructed above\ as
follows:%
\begin{equation}
\bar{F}_{s}(\text{ }\bar{\lambda}):=\bar{\lambda}  \label{4.12}
\end{equation}%
for some point $\bar{\lambda}\in \bar{B}_{1}(0).$

The solution of the fixed point equation (\ref{4.12})\ is, evidently,
completely equivalent to proving Theorem \ref{Th_4.1}, since the
corresponding vector $\bar{x}:=\xi (\bar{\lambda})\in M_{f},$ owing to
definition (\ref{4.11}), satisfies the following relationships:%
\begin{equation}
\bar{f}(\bar{x})=\bar{f}(\xi (\bar{\lambda}))=\xi (\bar{F}_{s}(\lambda
))\Rightarrow \xi (\bar{\lambda})=\bar{x}.  \label{4.13}
\end{equation}%
Thereby, the vector $\bar{x}:=\xi (\bar{\lambda})\in M_{f}$ solves fixed the
point problem (\ref{4.1}) for the compact mapping \ $\bar{f}:B\rightarrow B. 
$

To prove the existence of a solution to equation (\ref{4.12}), we will
construct the suitable \bigskip Banach spaces $E_{1}:=l_{1}(\mathbb{Z}_{+};%
\mathbb{R})\oplus \mathbb{R}$\ and $E_{2}:=l_{1}(\mathbb{Z}_{+};\mathbb{R})$%
\ and take the unit sphere $S_{1}(0)\subset E_{1},$ \ consisting of points $%
(\lambda ,\tau )\in E_{1},$\ for which $\ ||\lambda ||+|\tau |=1.$ The
mapping $\bar{F}_{s}:$ $\bar{B}_{1}(0)\rightarrow $ $\bar{B}_{1}(0),$
constructed above, one can extend upon the sphere $S_{1}(0)\subset E_{1},$
defining a mapping \ $f:E_{1}\supset S_{1}(0)\rightarrow \bar{S}%
_{1}(0)\subset E_{2}$ as 
\begin{equation}
f(\lambda ,\tau ):=\bar{F}_{s}(\lambda )  \label{4.14}
\end{equation}%
for any $(\lambda ,\tau )\in S_{1}(0)\subset E_{1}.$\ A suitable linear,
closed and surjective operator $\hat{a}:E_{1}\rightarrow E_{2}$ one can
define as 
\begin{equation}
\hat{a}\text{ }(\lambda ,\tau ):=\lambda  \label{4.15}
\end{equation}%
for all $(\lambda ,\tau )\in E_{1}.$ The resulting equation 
\begin{equation}
\hat{a}\text{ }(\lambda ,\tau )=f(\lambda ,\tau )  \label{4.16}
\end{equation}%
for $(\lambda ,\tau )\in S_{1}(0)\subset E_{1}$ exactly fits into the
conditions formulated in Theorem \ref{Th_3.1}, being simultaneously
equivalent to fixed point problem (\ref{4.12}) for the mapping $\bar{F}_{s}:$
$\bar{B}_{1}(0)\rightarrow $ $\bar{B}_{1}(0).$ Since $\dim Ker$ $\hat{a}=1,$
there exists the nonempty solution set ${\rm N}(\hat{a},f)\subset E_{1}$
of equation (\ref{4.16}). If a point $(\lambda _{\tau },\tau )\in {\rm
N}(\hat{a},f)\subset S_{1}(0),$ where $||\lambda _{\tau }||+|\tau |=1$ for
some value $|\tau |\leq 1,$ then the fixed point equality 
\begin{equation}
\bar{F}_{s}(\text{ }\lambda _{\tau }):=\lambda _{\tau }  \label{4.17}
\end{equation}%
holds for the value $\lambda _{\tau }\in \bar{B}_{1}(0)\subset l_{1}(\mathbb{%
Z}_{+};\mathbb{R}).$ Having denoted now $\lambda _{\tau }:=$\ \ $\bar{\lambda%
}\in \bar{B}_{1}(0),$ we get, owing to relationships (\ref{4.13}), the
corresponding solution to the fixed point problem for the compact mapping $%
\bar{f}:B\rightarrow B,$ thereby finishing the proof of the Leray-Schauder
theorem \ref{4.1}.
\end{proof}

There exist, evidently, many other interesting applications of the main
Theorem \ref{Th_3.1} in particular, proving the existence theorem for
diverse types of differential equations in Banach spaces with both fixed
boundary conditions and inclusions \cite{AE,Go2,Go1,Go3,Ge2,Ni}. These and
related research problems we plan to study in move detail in another paper.

\section{Acknowledgments}

The author is cordially indebted to the Abdus Salam International Centre for
Theoretical Physics, Trieste, Italy, for the kind hospitality during his
ICTP-2007 research scholarship. Special thanks are attributed to Professor 
Le Dung Tr\'{a}ng, Head of the Mathematical Department, for an invitation
to visit ICTP and creative atmosphere, owing to which the present work was
completed. The author is much appreciated to Professors Charles Chidume
(ICTP, Italy), Lech G\'{o}rniewicz (Juliusz Schauder Center, Torun, Poland)
and Anatoliy Plichko (Krakow Politechnical University, Poland) for very
useful discussions of the problems treated.

\end{document}